\documentclass{iau} 
\usepackage{graphicx}
\usepackage{natbib, color}
\usepackage[bookmarks=true, colorlinks=false, bookmarksopen]{hyperref}
\newcommand\power[2]{\ensuremath{#1\times10^{#2}}}

\newcommand\plotone[2]{\centering \leavevmode
  \includegraphics[width=#2\linewidth]{#1}}

\title[IAUS 317.~~Stellar populations in Illustris stellar halos] 
{Stellar populations of stellar halos: \\ Results from the Illustris simulation}

\author[Cook, Conroy, Pillepich, \& Hernquist]   
{B. A. Cook, C. Conroy, A. Pillepich, and L. Hernquist}

\affiliation{Harvard-Smithsonian Center for Astrophysics\\ 60 Garden
  St., Cambridge, MA 02138\\ Contact email: {\tt bcook@cfa.harvard.edu}}

\pubyear{2015}
\volume{317}  
\jname{The General Assembly of Galaxy Halos: Structure, Origin, and Evolution}
\editors{}
\begin{document}

\maketitle

\begin{abstract}
  The influence of both major and minor mergers is expected to
  significantly affect gradients of stellar ages and metallicities in
  the outskirts of galaxies. Measurements of observed gradients are
  beginning to reach large radii in galaxies, but a theoretical
  framework for connecting the findings to a picture of galactic
  build-up is still in its infancy. We analyze stellar populations of
  a statistically representative sample of quiescent galaxies over a
  wide mass range from the Illustris simulation. We measure
  metallicity and age profiles in the stellar halos of quiescent
  Illustris galaxies ranging in stellar mass from $10^{10}$ to
  $10^{12} M_\odot$, accounting for observational projection and
  luminosity-weighting effects. We find wide variance in stellar
  population gradients between galaxies of similar mass, with typical
  gradients agreeing with observed galaxies. We show that, at fixed
  mass, the fraction of stars born in-situ within galaxies is
  correlated with the metallicity gradient in the halo, confirming
  that stellar halos contain unique information about the build-up
  and merger histories of galaxies.
\end{abstract}

\firstsection 
\section{Introduction}

Stellar halos are diffuse regions of stars ubiquitously found
surrounding galaxies \citep[with a few notable exceptions,][]{VanDokkum2014b}. They are observed to extend to many times a
galaxy's effective radius \citep{Martinez-Delgado2010}, where
dynamical timescales are very long compared to the ages of their host
galaxies \citep{Eggen1962}. Detections of streams and tidal features
in integrated light images of the halos in the Milky
Way \citep{Helmi1999}, Andromeda
\citep{Ibata2001}, and external galaxies \citep{Martinez-Delgado2014}
imply that stellar halos result from mergers and accretion as galaxies
grow hierarchically, a picture anticipated by numerical simulations
\citep[e.g.,][]{Johnston1996,Bullock2005,Cooper2010,Pillepich2014}. The
long timescales at large radius can help preserve the information
content of each galaxy's merger history, which has been used as
motivation for observations of kinematics and populations of the
stellar halo \citep{Bell2008,Schlaufman2009}.

Stellar populations -- characterized by the metallicities and ages of
stars -- are clues to when and in what systems
stars originally formed \citep{Greene2015a}. The mass-metallicity
relation in galaxies \citep{Tremonti2004}, combined with hierarchical
accretion, implies that metallicity gradients should indicate the
relative contributions of more-or-less massive progenitor systems to
particular regions of a galaxy and its halo. Different merger
histories should leave distinct imprints on observed stellar
population gradients \citep{Hirschmann2015}, although stars formed
\textit{in-situ}, or within the galaxy in which they now reside, may
still be significant in the inner regions of the halo
\citep{Font2011}. For massive, \textit{early-type} galaxies (usually
red, quiescent, and elliptical), the two-phase formation scenario
outlines the contributions of mergers and in-situ star formation to
the formation of a metallicity gradient
\citep{Spolaor2010b,Pastorello2014}. The early phase of dissipative
collapse leads to an steep, negative gradient from in-situ stars,
while the later phase involving accretion of stars from smaller
satellite galaxies tends to flatten these gradients
\citep{Kobayashi2004}.

Observations of stellar population gradients are beginning to
accumulate for large samples of galaxies
\citep{Spolaor2010b,Pastorello2014} and are reaching continually
larger radii through integral-field spectroscopy
\citep{Delgado2015,Greene2015a}. Observations find a wide variety of
gradients between galaxies with similar masses and morphologies, which
is to be expected if particular merger histories shape the gradients
in a stochastic way. Unfortunately, the observational need for
detailed predictions connecting gradients to merger histories have so
far outpaced numerical simulations. Due to limitations in computing
resources, hydrodynamical simulations have so far only resolved
stellar halo populations around individual galaxies
\citep{Abadi2006,Hirschmann2015,Cooper2015a}, a method which lacks the
statistical power required to replicate the observed halo-to-halo
variations. Large samples of galaxy halos have been produced in N-body
simulations, some of which include semi-analytic models and
stellar-tagging techniques \citep{Cooper2010}, but these do not
include baryonic physics effects, which may shape the dark matter
distribution, and must rely on complicated fitting functions to
generate ``stellar particles'' with realistic orbital properties
\citep{Bailin2014}.

In this work, we measure stellar population gradients in a sample of
quiescent galaxies from the hydrodynamical cosmological simulation
Illustris. With the statistical power of Illustris' large cosmological
volume and its self-consistent model for star and galaxy formation, we
are able to match the observed galaxy-to-galaxy variance in gradients
and show that these variations are indeed reflective of different
galactic merger histories.
\section{Methods}
\subsection{The Simulations}
The Illustris simulations
\citep{Vogelsberger2014,Genel2014a,Nelson2015} are a suite of
N-body+hydrodynamical cosmological simulations (106.5 Mpc on a side),
run at multiple resolutions with the adaptive mesh code \textit{AREPO}
\citep{Springel2010,Vogelsberger2013}. The simulations model key
physical processes for the formation of galaxies, including stellar
formation, evolution, and feedback, chemical enrichment, radiative
cooling, supermassive black hole growth, and feedback from AGN. The
highest resolution run (Illustris-1, hereafter simply Illustris) has a
mass resolution of $m_{DM} = \power{6.26}{6} M_\odot$ and
$m_{baryon} \sim \power{1.26}{6} M_\odot$ for the dark matter and
baryonic components, respectively. At $z=0$ gravitational forces for
stellar particles are resolved to a softening length of 0.7
kpc. Illustris was run from $z=127$ to $z=0$ using $\Lambda$CDM
cosmological paramters consistent with \textit{WMAP9}
\citep[$\Omega_\Lambda = 0.7274, \Omega_m = 0.2726, h =
  0.704$,][]{Hinshaw2013}.

At $z=0$, the Illustris volume contains more than \power{4}{4}
well-resolved galaxies \citep{Vogelsberger2014}, with a reasonable
diversity of morphologies and colors, including early-type and
late-type galaxies \citep{Torrey2015a}. The most massive central
galaxies \citep[as identified by the \textit{FOF} and \textit{SUBFIND}
  algorithms,][]{Springel2001,Dolag2009} reproduced in the simulation
have stellar masses within their stellar half-mass radii of $M_* \sim
\power{1-2}{12} M_\odot$. The simulation reproduces the observed
$z=0$ mass-metallicity relation in galaxies \citep{Vogelsberger2014a},
and a reasonable relation between galaxy mass and stellar ages once
luminosity weighting is properly taken into account.

\subsection{The Quiescent Galaxy Sample}

Our goal to reproduce measurements in the outer regions of early-type
galaxies motivates the selection criteria of our sample. We select
central galaxies (which are not satellites/subhalos of more massive
parents) with stellar masses $M_* \geq 10^{10}
M_\odot$$^{\footnote{}}$\footnotetext{Throughout the paper, $M_*$ denotes the total mass
  in stars within the stellar half-mass radius ($R_{half}$) after
  removing the contributions of gravitationally-bound satellites
  identified by \textit{SUBFIND}.}. Each galaxy in this mass range is
resolved with at least a few hundred star particles beyond 8 effective
radii, ensuring that we constrain gradients all the way throughout the
low-density outer regions and up to 10 times the effective radius.

In this work, we examine the properties of simulated
\textit{quiescent} galaxies, in comparison to observed
\textit{early-type} galaxies. Of all Illustris galaxies with $M_* >
10^{10} M_\odot$, we select the 352 quiescent galaxies with specific
star formation rates SSFR $\leq 10^{-11.5}$ yr$^{-1}$ within twice the
stellar half-mass radius. See Cook et al. (In Prep.) for details.

\subsection{Fitting Stellar Population Gradients}

We simulate observational projection effects by projecting the radii
of star particles from the center of their host galaxies against a
random line-of-sight. We also account for observational biases by
weighting the contribution of each star particle relative to its
V-band lumionsity$^{\footnote{}}$\footnotetext{Star particles in each galaxy are assigned
  luminosities in several common observational bands using single-age
  stellar population SED templates \citep{Torrey2015a}.}.
Star particles identified by \textit{SUBFIND} as bound to smaller
subhalos or satellites are not included in the analysis.

Using these observational considerations, we measure the 2-D
azimuthally-averaged values of stellar metallicity and age in five
logarithmically-spaced radial bins over a chosen projected radius
range. We focus on three particular ranges in units of the V-band
effective radius ($R_e\,^{\footnote{}}$\footnotetext{Each galaxy's effective radius is the
  projected radius within which one-half of its total V-band
  luminosity is located. This is calculated as an average over 100
  random lines of sight.}), which we label: the inner galaxy (0.1 - 1
$R_e$), the outer galaxy (1 - 2 $R_e$), and the stellar halo (2 - 10
$R_e$). We then calculate a logarithmic gradient by fitting a line to
these averages, equally weighting each radius bin:
\begin{equation}
  f(r) = f(R_e) + \nabla_f \log_{10} (r / R_e),
\end{equation}
with $f$ the mean log-metallicity ([Z/H]) or the
mean age (in Gyrs) in each bin.

\section{Results and Implications}
\subsection{Comparisons to Observed Gradients}
\begin{figure*}[bt]
  \plotone{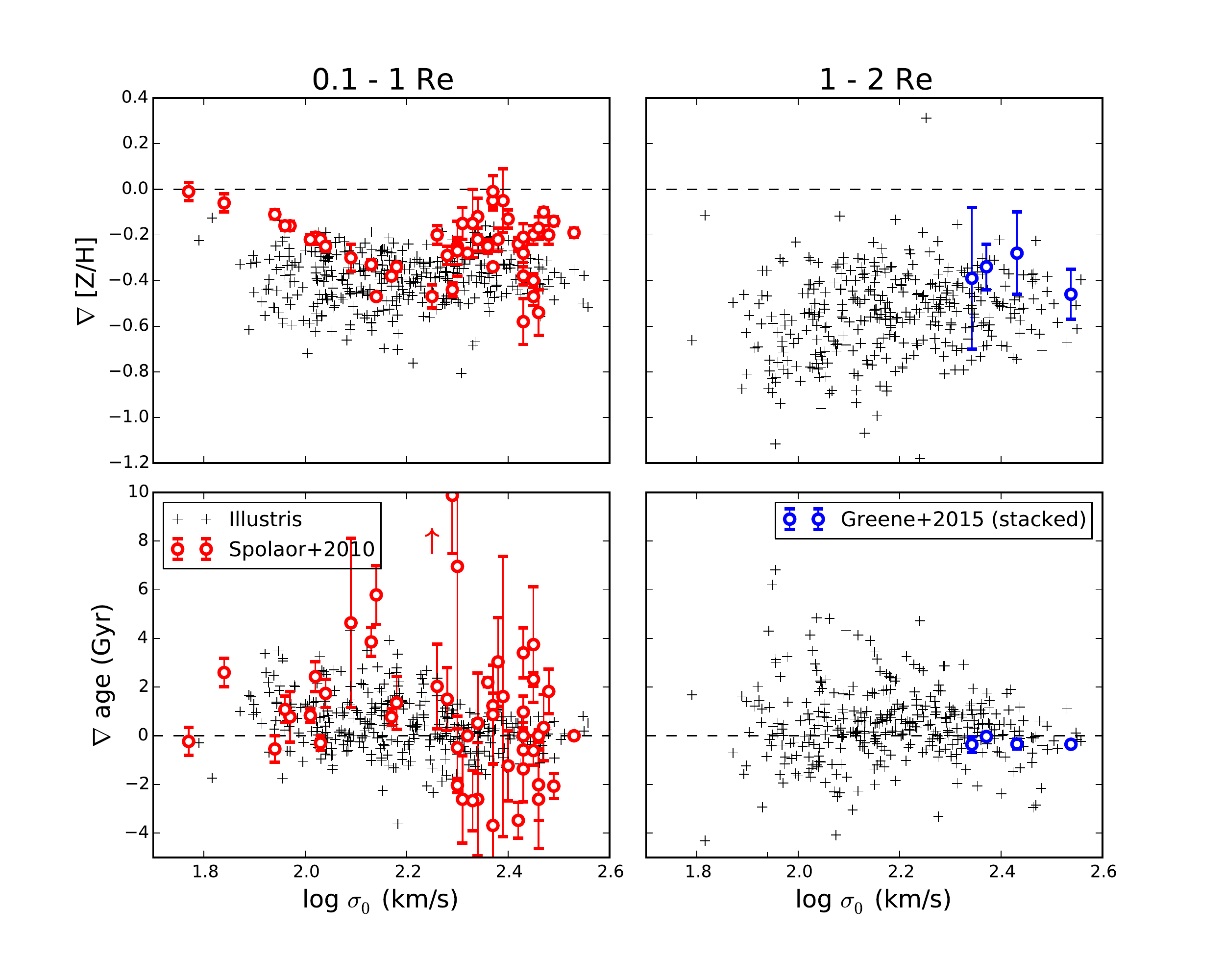}{0.98}
  \caption{Measurements of metallicity (\textit{top}) and age
    (\textit{bottom}) gradients in Illustris quiescent galaxies (black
    cross marks). These are calculated in the inner (0.1 - 1 $R_e$,
    \textit{left}) and outer (1 - 2 $Re$, \textit{right}) galaxy
    ranges, and include projection along a random line of sight and
    luminosity-weighting. We compare to the measurements of
    \citet{Spolaor2010b} and \citet{Greene2015a} in the inner and
    outer galaxy, respectively.}
  \label{fig:grads_compare}
\end{figure*}

The measured metallicity and age gradients in the inner and outer
galaxy ranges ($< 2 R_e$) are shown as a function of central velocity dispersion
($\sigma_0$, within $\frac{1}{8} R_e$) in
Fig.~\ref{fig:grads_compare}. Observations can constrain gradients in
these radius ranges for large samples of galaxies
\citep{Spolaor2010b,Pastorello2014,Greene2015a,Delgado2015}, but so
far there are only a few individual cases for comparison in the
stellar halo ($> 2R_e$) range. We compare our measurements to
observations of individual galaxies from \citet{Spolaor2010b} (inner
galaxy) and stacked measurements by \citet{Greene2015a} (outer
galaxy).

The typical values of our metallicity and age gradients agree with
observations: metallicity gradients are negative -- outer regions are
more metal poor -- while age gradients are roughly flat -- inner and
outer regions have similar average ages. Of particular importance is
the fact that our measurements also reproduce the observed scatter in
gradients between galaxies of similar masses. Illustris, with its
large statistical sampling of galaxies, is uniquely able to replicate
this galaxy-galaxy variance, whereas more individualized simulations
cannot replicate the wide variety of observed gradients.

\subsection{Relating Metallicity Gradients to Merger Properties}

Here, we study whether the scatter in stellar population gradients
among Illustris galaxies at fixed mass -- which matches the scatter
observed in early-type galaxies -- can be connected to the particular
merger histories of each galaxy. We quantify the contribution of
mergers to a galaxy's build-up in terms of the \textit{in-situ
  fraction}: the mass fraction of a galaxy's stars which were formed
within the galaxy or its main progenitor branch. Galaxies with low
in-situ fractions were primarily built-up from mergers and accretion of
smaller systems, while galaxies with high in-situ fractions have had
little influence from mergers.

In the hierarchical model of galaxy formation, galaxies grow
continually through mergers with increasingly massive neighbors. This
effect, reproduced in many cosmological simulations and seen clearly
in Illustris \citep{Rodriguez-Gomez2015a}, results in a strong trend
towards lower in-situ fractions in more massive galaxies. To
disentangle this effect from the individual histories, we can focus
our analysis to small ranges in stellar mass. Over a sufficiently
narrow window, mass differences are small enough that the in-situ
fraction should be driven primarily by the particular merger history of
each galaxy.

\begin{figure*}[bt]
  \plotone{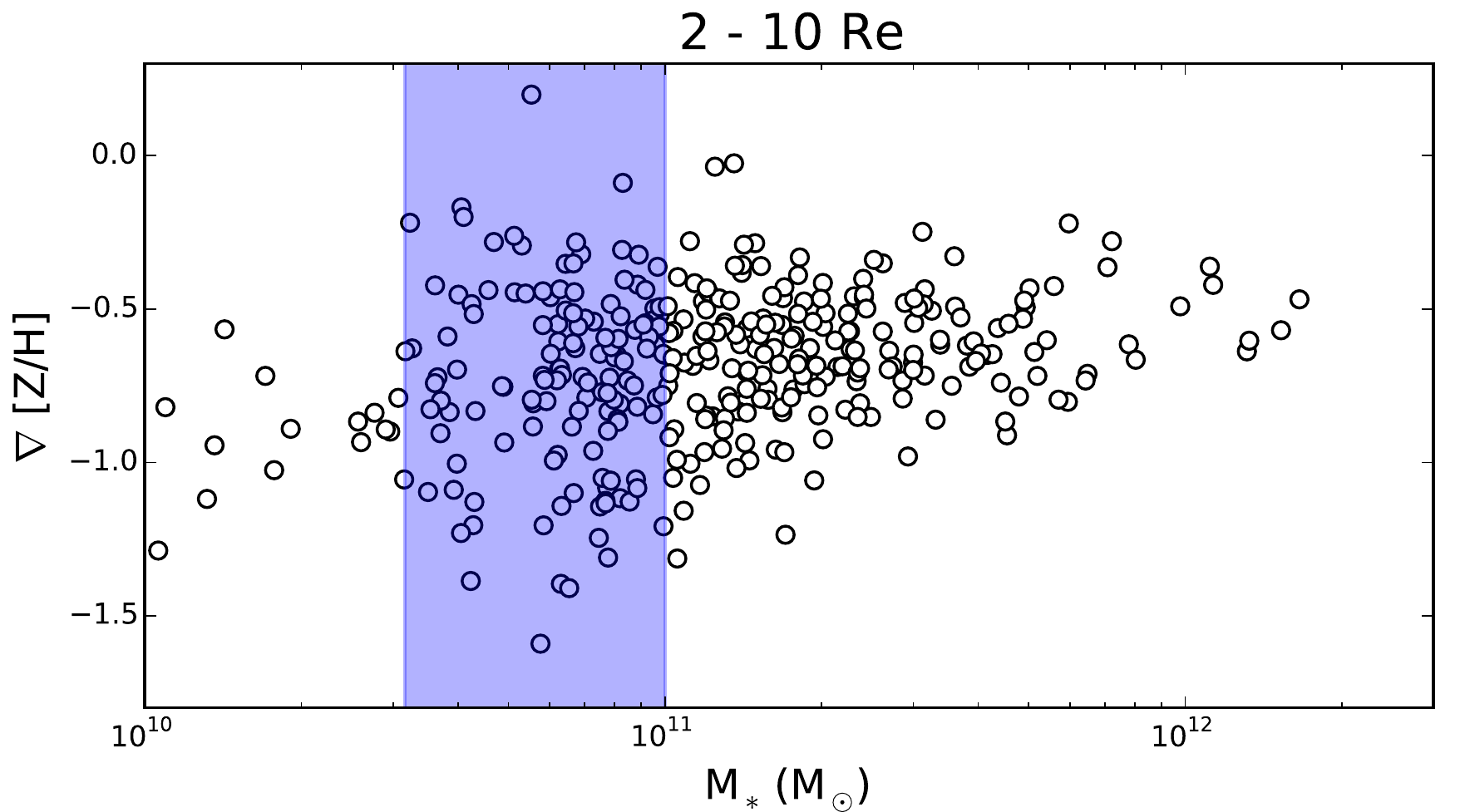}{0.8}
  \plotone{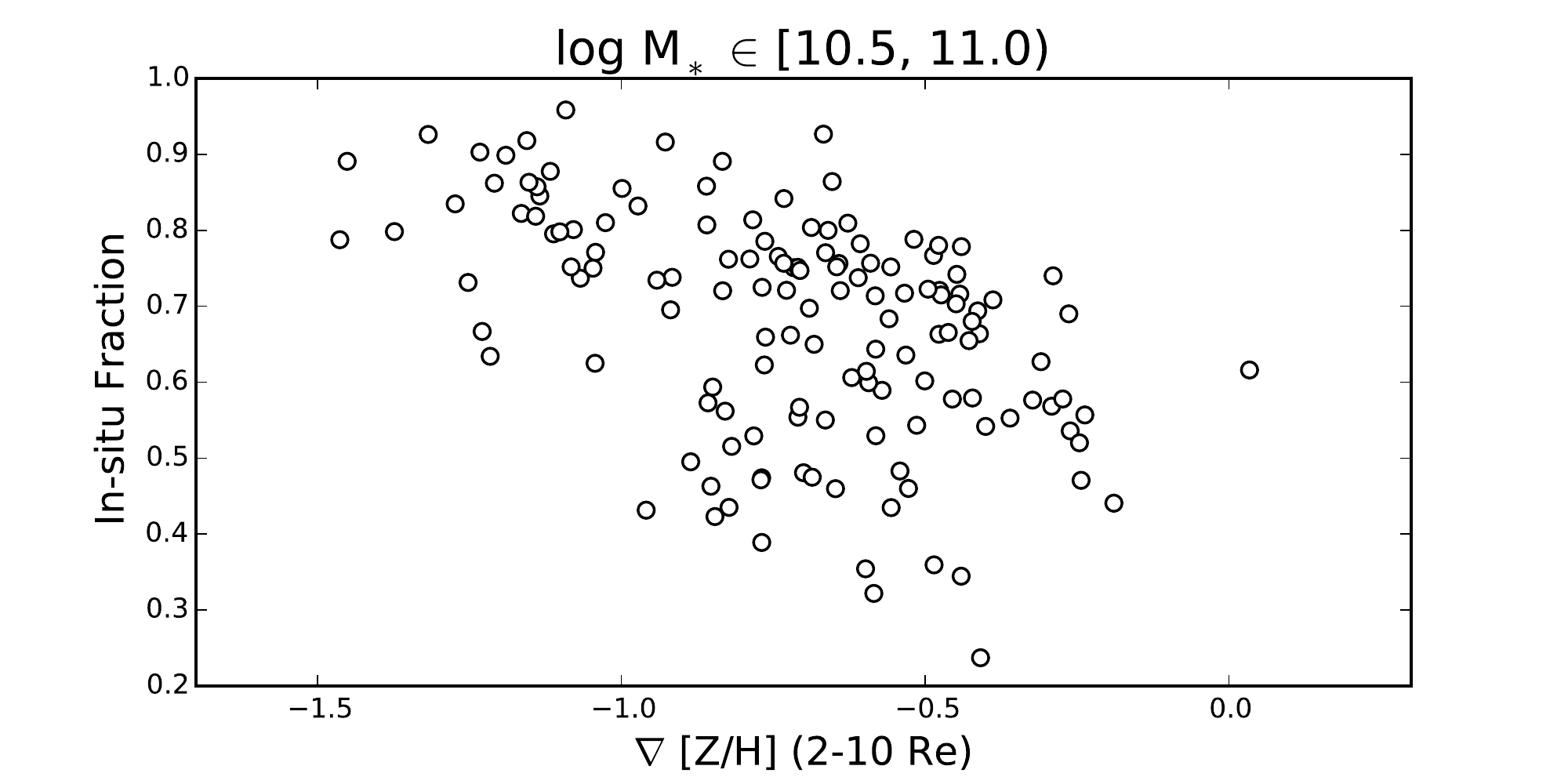}{0.8}
  \caption{\textit{Top}: The metallicity gradient in the stellar halo
    (2 - 10 $R_e$) of the quiescent Illustris sample, versus galaxy
    stellar mass. As at smaller radii, the metallicity gradient shows
    wide galaxy-galaxy variance within narrow mass ranges, such as
    between $10^{10.5}$ and $10^{11} M_\odot$ (shaded
    region). \textit{Bottom}: The correlation between galaxy in-situ
    fraction and stellar halo metallicity gradient, for galaxies
    between in the mass range given above. Galaxies with lower
    in-situ fractions (mergers more significant in growth) have
    flatter metallicity gradients.}
  \label{fig:correlation}
\end{figure*}

In Fig.~\ref{fig:correlation}, we show the metallicity gradient in the
stellar halo (2 - 10 $R_e$). Galaxies have overall negative
metallicity gradients and there is a large scatter at fixed mass, just
as in the inner and outer galaxy regions. When selecting galaxies in
the narrow mass window from $10^{10.5}$ to $10^{11} M_\odot$ and
comparing to the in-situ fraction, we see that there is a
correlation. Low in-situ fractions are associated with flatter
metallicity gradients in the stellar halo, while galaxies with higher
in-situ fractions have steeper halo gradients.

This result suggests that much of the scatter in observed metallicity
gradients can be explained through a variety of galactic merger
histories. The process of building-up a galaxy through mergers appears
to flatten the metallicity gradient. Measuring gradients into the
outer stellar halo should provide clues about the relative importance
of mergers in a particular early-type galaxy's evolution \citep[see
  also,][]{Pillepich2014}.

We note that this correlation is significant only in the stellar halo
regions (2 - 10 $R_e$) of our simulated galaxies. Metallicity gradients
in the inner and outer galaxy regions ($< 2 R_e$) show no association
with the in-situ fraction. This emphasizes the importance of measuring
gradients into the outer stellar halo (well beyond $2 R_e$), where the
significantly longer dynamical timescales result in more lasting
imprints from mergers. Stellar population gradients in the halo retain
the information content of merger histories; it is still unclear what
leads to the wide scatter in gradients in the interior regions.

\acknowledgments
This material is based upon work supported by the NSF Graduate
Research Fellowship Program under Grant No. DGE1144152.

\bibliography{StellarHalos}

\end{document}